\documentclass[apj]{emulateapj}
\usepackage{amsmath}

\makeatletter

\newenvironment{inlinefigure}{%
\def\@captype{figure}%
\noindent\begin{minipage}{0.999\linewidth}\begin{center}}
{\end{center}\end{minipage}\smallskip}
\makeatother

\begin{document}

\title{A Submillimeter Perspective on the GOODS Fields---II.
The high radio power population in the GOODS-N\altaffilmark{1,2,3,4}}

\author{
A.~J.~Barger\altaffilmark{5,6,7}, 
L.~L.~Cowie\altaffilmark{7},
F.~N.~Owen\altaffilmark{8},
L.-Y.~Hsu\altaffilmark{7},
W.-H.~Wang\altaffilmark{9}
}

\altaffiltext{1}{The James Clerk Maxwell Telescope is operated by the 
East Asian Observatory on behalf of The National Astronomical
Observatory of Japan, Academia Sinica Institute of Astronomy and
Astrophysics, the Korea Astronomy and Space Science Institute,
the National Astronomical Observatories of China and the Chinese
Academy of Sciences (Grant No. XDB09000000), with additional
funding support from the Science and Technology Facilities Council
of the United Kingdom and participating universities in the United
Kingdom and Canada.}
\altaffiltext{2}{The National Radio Astronomy Observatory is a facility of 
the National Science Foundation operated under cooperative agreement by 
Associated Universities, Inc.}
\altaffiltext{3}{The Submillimeter Array is a joint project between the
Smithsonian Astrophysical Observatory and the Academia Sinica Institute
of Astronomy and Astrophysics and is funded by the Smithsonian Institution
and the Academia Sinica.}
\altaffiltext{4}{The W.~M.~Keck Observatory is operated as a scientific
partnership among the California Institute of Technology, the University
of California, and NASA, and was made possible by the generous financial
support of the W.~M.~Keck Foundation.}
\altaffiltext{5}{Department of Astronomy, University of Wisconsin-Madison,
475 N. Charter Street, Madison, WI 53706, USA}
\altaffiltext{6}{Department of Physics and Astronomy, University of Hawaii,
2505 Correa Road, Honolulu, HI 96822, USA}
\altaffiltext{7}{Institute for Astronomy, University of Hawaii,
2680 Woodlawn Drive, Honolulu, HI 96822, USA}
\altaffiltext{8}{National Radio Astronomy Observatory, P.O. Box O, 
Socorro, NM 87801, USA}
\altaffiltext{9}{Academia Sinica Institute of Astronomy and Astrophysics, 
P.O. Box 23-141, Taipei 10617, Taiwan}

\slugcomment{ApJ, 835, 95 (2017)}

\begin{abstract}
We use ultradeep 20\,cm data from the Karl G. Jansky Very Large Array
and 850\,$\mu$m data from SCUBA-2 and the Submillimeter Array  
of an 124~arcmin$^2$ region of the {\em Chandra\/} Deep Field-north
to analyze the high radio power ($P_{20\,\rm cm}>10^{31}$~erg~s$^{-1}$~Hz$^{-1}$) 
population. We find that 20 ($42\pm9$\%) of the spectroscopically
identified $z>0.8$ sources have consistent star formation
rates (SFRs) inferred from both submillimeter and radio observations, 
while the remaining sources have lower (mostly undetected) 
submillimeter fluxes, suggesting that active galactic nucleus (AGN) 
activity dominates the radio power in these sources.
We develop a classification scheme based on the ratio of submillimeter flux to radio 
power versus radio power and find that it agrees with AGN and 
star-forming galaxy classifications from Very Long Baseline Interferometry.
Our results provide support for an extremely rapid drop in the number of high SFR
galaxies above about a thousand solar masses per year (Kroupa initial mass function)
and for the locally determined relation between X-ray luminosity and radio
power for star-forming galaxies applying at high redshifts and high radio powers.
We measure far-infrared (FIR) luminosities 
and find that some AGNs lie on the FIR-radio correlation, while others scatter below.
The AGNs that lie on the correlation appear to do so based on their emission from 
the AGN torus. 
We measure a median radio size of $1\farcs0\pm0.3$ for the star-forming galaxies.
The radio sizes of the star-forming galaxies are generally larger than those of the AGNs.

\end{abstract}


\section{Introduction}
\label{secintro}

The local 20\,cm population may be roughly divided into primarily star-forming galaxies 
at low radio power
and primarily active galactic nucleus (AGN) dominated galaxies at high radio power 
(e.g., Condon 1989; Sadler et al.\ 2002; Mauch \& Sadler 2007; Best \& Heckman 2012).
Such separations by galaxy type are mostly based on spectroscopic identifications.
However, as we move to higher redshifts, many of the radio sources become too faint for
spectroscopic identifications. For example, in Barger et al.\ (2014), we found that an extended
tail of microJansky radio sources with faint near-infrared (NIR) counterparts ($K_S>22.5$) 
make up $\sim30$\% of the 20~cm population in the extremely deep 
(11.5~$\mu$Jy at $5\,\sigma$) Karl G. Jansky Very Large Array (VLA) image 
(F. Owen 2017, in preparation) of the {\em Chandra\/} Deep Field-north 
(CDF-N; Alexander et al.\ 2003).
To complicate matters, at high redshifts there are many more galaxies
with extreme star formation rates (SFRs). Thus, the high radio power population
begins to contain a significant fraction of star-forming galaxies, as well as continuing to be 
populated by numerous AGNs, resulting in  a complex function with redshift of these 
competing processes (e.g., Cowie et al.\ 2004a).

In this paper, we develop a method to separate star-forming galaxies---galaxies 
where the SFRs inferred from the submillimeter observations are 
consistent with those inferred from the radio power---from AGNs in the high 
radio power population using a combination of ultradeep 20\,cm and 
850\,$\mu$m observations of the CDF-N. 
Long-wavelength submillimeter
observations provide a clean discrimination between AGNs, whose spectral
energy distributions (SEDs) drop extremely rapidly above a rest-frame wavelength 
of about $100\,\mu$m (e.g., Mullaney et al.\ 2011), and
galaxies dominated by star formation, whose graybody emission extends
smoothly to the submillimeter. Both radio-quiet AGNs and star-forming galaxies 
can lie on the well-known FIR ($8-1000\,\mu$m)-radio correlation; however, because of 
their very different long-wavelength properties, they can be distinguished 
using the present technique.
We then compare our classifications with Very Long Baseline Interferometry (VLBI)
classifications and with the X-ray properties of the sources.

Although submillimeter observations provide a valuable means of picking 
out extreme star formers in the high radio power population, observing large 
areas to any significant depth in the submillimeter is very difficult. 
Thus, we conclude by looking for a simpler diagnostic for 
separating star-forming galaxies from AGNs based on the radio data alone,
namely, the radio size.

Our primary science goals are (1) to look for evidence of a
characteristic maximum SFR for galaxies, as proposed 
by Karim et al.\ (2013) and Barger et al.\ (2014), (2) to determine whether
locally determined relations between X-ray luminosity and radio power for 
star-forming galaxies apply at high redshifts and high radio powers,
(3) to find whether AGNs in the high-redshift, high radio power population lie 
on the FIR-radio correlation (de Jong et al.\ 1985; Helou et al.\ 1985), as has
been observed locally (Condon 1992; Mori{\'c} et al.\ 2010; Wong et al.\ 2016),
and (4) to examine whether any AGNs that lie on the correlation are
there because star formation in the host galaxy is driving the observed radio 
emission, as proposed by Condon et al.\ (2013).

In Section~\ref{secdata}, we present the radio, submillimeter, X-ray, and
spectroscopic and photometric redshift data sets that we use in our analysis.
In Section~\ref{secsfgs}, we develop our classification method using the
ratio of submillimeter flux to radio power and we
compare our classifications of the high radio power sources with those made 
using radio-excess measurements, VLBI, and X-rays.
In Section~\ref{seccorr}, we measure the FIR luminosities for the high radio power
sources and determine their locations on a FIR luminosity versus radio power plot.
We then construct
SEDs for the star-forming galaxies and for the AGNs separately 
and determine whether they are consistent with the idea that star formation in the 
host galaxies is responsible for putting some AGNs on the FIR-radio correlation.
Finally, in Section~\ref{secsizes}, we explore whether the separation between
star-forming galaxies and AGNs could be done from radio sizes alone.

We assume the Wilkinson Microwave
Anisotropy Probe cosmology of $H_0=70.5$\,km\,s$^{-1}$\,Mpc$^{-1}$,
$\Omega_{\rm M}=0.27$, and $\Omega_\Lambda=0.73$ 
(Larson et al.\ 2011) throughout.

\section{Data}
\label{secdata}
\subsection{Radio and Submillimeter}
We use ultradeep radio and submillimeter observations
of the GOODS-N/CDF-N. 
The radio data are the ultradeep 20~cm image obtained in a 40\,hr integration 
using the upgraded VLA in $A$-configuration (F. Owen 2017, in preparation). 
The image covers a $40'$ diameter with an effective resolution of $1\farcs8$.
The highest sensitivity region is about $9'$ in radius.
The image probes to a $1\,\sigma$ limit of 2.3\,$\mu$Jy, 
making it the deepest radio image currently available at this wavelength.
The absolute radio positions are known to $0\farcs1$--$0\farcs2$ rms.
We detect 445 radio sources above a $5\,\sigma$ threshold in the 
selected 124~arcmin$^2$ SCUBA-2 area. 
Of these, 210 are spatially resolved in the radio image. We measured 
sizes for these resolved sources and upper limits for the remaining sources.

The submillimeter data consist of an 101~hr
850\,$\mu$m image that we obtained with SCUBA-2
(Holland et al.\ 2013) on the 15\,m James Clerk Maxwell Telescope, 
together with follow-up high-resolution observations that we made with the 
Submillimeter Array (SMA; Ho et al.\ 2004). 
The center of the SCUBA-2 image has an rms depth of 0.28\,mJy.
In this paper, we consider an area of 124\,arcmin$^2$ where
the rms is less than 0.57\,mJy (twice the central noise). 
For a detailed description of the reduction and calibration of SCUBA-2 data
in general,
we refer the reader to Chapin et al.\ (2013) and Dempsey et al.\ (2013).
For specific details on the SCUBA-2 CDF-N image data reduction and calibration, 
we refer the reader to Chen et al.\ (2013) and Cowie et al. (2016).

We formed a matched filter image by weighting
the SCUBA-2 image with the point spread function (PSF). This provides an optimal
estimate of the flux at any position provided that, as expected,
the sources are small compared with the beam full width half
maximum (FWHM) of $14''$ at 850\,$\mu$m. We used a wider filter to subtract
variable backgrounds so that the average measured
flux at random positions in the image equals zero.
We detect 115 sources at $>4\,\sigma$ in the 124~arcmin$^2$ area.
Of these, 41 have fluxes above 2.85\,mJy, which is the flux threshold
corresponding to a $>5\,\sigma$ detection throughout the area. 

Our SMA follow-up observations, together with archival data, give detections and
high-precision subarcsecond positions for 22 submillimeter sources
in this area, including all eight SCUBA-2 sources with fluxes above 6\,mJy. 
Catalogs and further details on the SCUBA-2 image and the SMA data
may be found in  Cowie et al.\ (2016). Barger et al.\ (2014, 2015) used a
substantial subset of the CDF-N SCUBA-2 data, as well as the SMA data, 
in their analyses.

In measuring the 850\,$\mu$m fluxes for the radio sources without
SMA counterparts, we first removed all of the detected SMA sources
from the matched filter  SCUBA-2 image using a PSF based 
on the observed calibrators. This left residual images from which we measured the 
850\,$\mu$m fluxes (whether positive or negative) and statistical errors at the radio positions.
This procedure minimizes contamination by brighter submillimeter sources in the field.

\subsection{X-Ray}
\label{xraydata}
The X-ray data are the 2\,Ms X-ray image of Alexander et al.\ (2003), which they
aligned with the Richards (2000) radio image. Near the aim point, the X-ray catalog
reaches a limiting flux of 
$f_{0.5-2~\rm keV}\approx 1.5\times 10^{-17}$\,erg\,cm$^{-2}$\,s$^{-1}$.
Matching X-ray counterparts from the Alexander et al.\ (2003) catalogs to the
radio sources is not critically dependent on the choice of match radius, as can be
seen from Figure~7 of Alexander et al.\ (2003), which shows the positional
offset between the X-ray and radio sources versus off-axis angle.  Following
Barger et al.\ (2007), we use a $1\farcs5$ search radius.
Of the radio sources in the CDF-N sample, 142 have X-ray counterparts.
(There are a further 138 X-ray sources in the region that do not have radio counterparts.)
When a radio source does not have an X-ray counterpart in the catalog,
we measure the X-ray counts at the radio position from the X-ray image using 
a $3''$ diameter and convert them to fluxes assuming a fixed photon index 
of $\Gamma=1.8$. For all the sources, we compute the rest-frame $2-8$\,keV 
luminosities, $L_X$, from the $0.5-2$\,keV fluxes with no absorption correction and 
$\Gamma=1.8$ using
\begin{equation}
L_X = 4\pi d_L^2 f_{0.5-2\,{\rm keV}} ((1+z)/4)^{\Gamma-2}~{\rm erg~s^{-1}} \,.
\end{equation}
We take $L_X>10^{44}$\,erg\,s$^{-1}$ as the threshold for a source to be
classified as an X-ray quasar.

\subsection{Redshifts}
There is a large amount of ancillary data on the CDF-N.
However, the most important information for this analysis is the spectroscopic 
data. Here we draw on a new compilation of known spectroscopic
redshifts in the region (A. Barger et al.\ 2017, in preparation;
see also Cohen et al.\ 2000; Cowie et al.\ 2004b, 2016; Swinbank et al.\ 2004; 
Wirth et al.\ 2004, 2015; Chapman et al.\ 2005; Reddy et al.\ 2006; 
Trouille et al.\ 2008; Barger et al.\ 2008; Cooper et al.\ 2011), the bulk of which 
come from Keck spectroscopy with DEIMOS, LRIS, and MOSFIRE. The compilation
includes redshifts based on our own analysis of the HST grism
data on the GOODS-N (PI: B Weiner; Momcheva et al. 2016), but this
only adds one redshift which is not identified otherwise.

Of the 445 radio sources in the 124~arcmin$^2$ area, 
343 (77$\%$) have spectroscopic redshifts. 
These include a small number of CO spectroscopic redshifts from 
Daddi et al.\ (2009a,b), Bothwell et al.\ (2013), and Walter et al.\ (2012).
We use photometric redshifts
from Rafferty et al.\ (2011) to augment the spectroscopic
redshifts, which raises the redshift identifications to 392. The
remaining 53 radio sources are very faint in the optical and
NIR with median $z$-band magnitudes of 26.2 and median $K_S$-band 
magnitudes of 23.1.

\subsection{High Radio Power Sources}
For all of the sources with either spectroscopic or photometric
redshifts, we compute the rest-frame radio power using the equation
\begin{equation}
P_{20\,\rm cm}=4\pi {d_L}^2 f_{20\,\rm cm} 10^{-29}
(1+z)^{\alpha - 1}~{\rm erg~s^{-1}~Hz^{-1}} \,,
\label{eqradio}
\end{equation}
where $d_L$ is the luminosity distance (cm)  and $f_{20\,\rm cm}$ is the
20\,cm flux in units of $\mu$Jy.
This equation assumes $S_\nu\propto \nu^{-\alpha}$, where we adopt a radio spectral 
index of $\alpha=0.8$ (Condon 1992; Ibar et al.\ 2010). The choice of $\alpha$
may not be appropriate for AGNs and also may be problematic
for high-redshift sources. However, it should be accurate
enough for the present purposes. 

There are 46 high radio power sources (defined here as
$P_{20~\rm cm}>10^{31}$~erg~s$^{-1}$~Hz$^{-1}$) with spectroscopic 
or photometric redshifts
in the 124~arcmin$^2$ area. All of these lie
at $z>0.8$. The lowest redshift is $z=0.847$, and the highest redshift
is $z=5.18$  (see Figure~\ref{new_radio_z}).
We return to the sources with no redshift identifications at the end of Section~\ref{secsfgs}.

\vskip 0.4cm
\begin{inlinefigure}
\includegraphics[angle=0,width=3.2in]{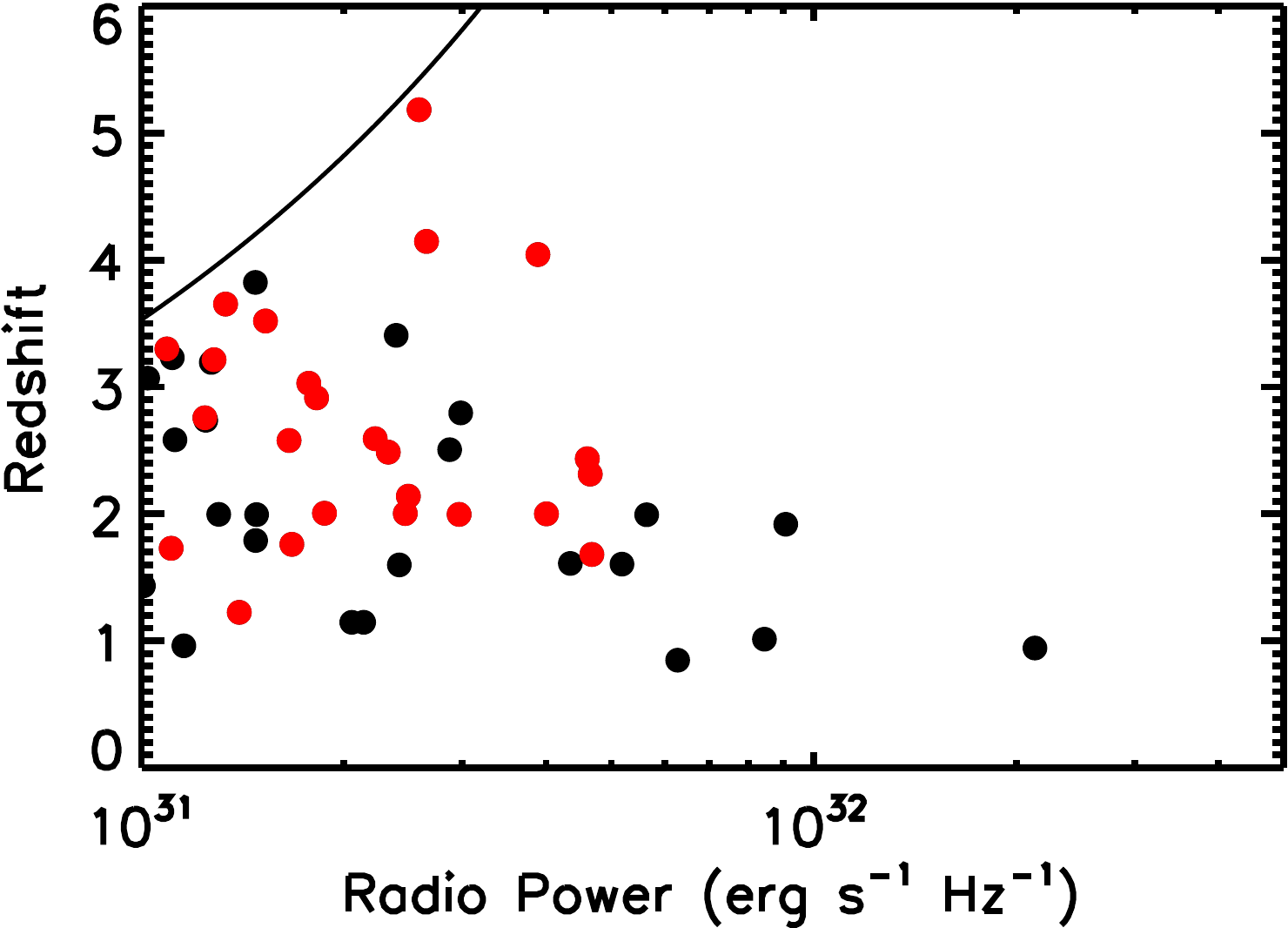}
\caption{Redshift vs. radio power. The detection
threshold for the radio sample is shown with the
black curve. Red circles show sources detected
above the $3\,\sigma$ level at 850$\,\mu$m, while black
circles show sources not detected at this level.
\label{new_radio_z}
}
\end{inlinefigure}

In Figure~\ref{new_radio_z}, we show the distribution of redshifts versus
radio power, distinguishing sources with submillimeter detections ($>3\sigma$;
red circles) from those without (black circles). Most of the sources lie
in the redshift range $z=1-3$.  The spectroscopic and photometric redshift
measurements may introduce biases in the identifications. In particular,
the highest redshift ($z>4$) sources are all based on CO measurements; thus,
all are submillimeter-detected sources. However, this does not affect
our analysis, since we compare only identified sources, and such a 
comparison does not depend on the completeness of the sample.

\subsection{The FIR-radio Correlation}
The local FIR-radio correlation (de Jong et al.\ 1985; Helou et al.\ 1985)
is well characterized by a linear fit over five orders of magnitude. It is
parameterized by the quantity
\begin{equation}
q = \log \left(\frac{L_{8-1000~\mu {\rm m}}}{3.75\times 10^{12}~{\rm erg~s^{-1}}} \right) - \log \left(\frac{P_{\rm 20~cm}}{\rm erg~s^{-1}~Hz^{-1}} \right) \,.
\label{firradio}
\end{equation}
The average local value is 
$q=2.52\pm0.01$ for an $8-1000\,\mu$m luminosity (Yun et al\ 2001;
this is after applying their correction of 1.5 from their measured
$42.5-122.5\,\mu {\rm m}$ luminosity; see also Bell 2003). 
This is very similar to the 
values measured at high redshifts from submillimeter samples with 
accurate positions from submillimeter interferometric observations 
(e.g., average $q=2.51\pm0.01$, Barger et al.\ 2012; 
median $q=2.56\pm0.05$, Thomson et al.\ 2014).

\section{Star-Forming Galaxies versus AGNs}
\label{secsfgs}
Barger et al.\ (2014) noticed in their plot of 850\,$\mu$m flux 
(their 850\,$\mu$m rms error was $<1.5$~mJy) 
versus radio power that submillimeter-detected ($>4\,\sigma$) radio sources 
(submillimeter-bright) bifurcate from other radio sources (submillimeter-blank)
above $P_{20~\rm cm}\approx 10^{31}$~erg~s$^{-1}$~Hz$^{-1}$. Based on a small 
number (5) of such sources that also had high-resolution radio observations 
(Chapman et al.\ 2004; Muxlow et al.\ 2005;
Momjian et al.\ 2010; Guidetti et al.\ 2013), Barger et al.\ found that the 
submillimeter-bright sources appeared to be extended and star-formation dominated, 
while the submillimeter-blank sources appeared to be compact.

Here we explore this observed bifurcation further using our deeper submillimeter data.
In Figure~\ref{barthel}, we plot 850~$\mu$m flux versus radio power for the high radio 
power sources with spectroscopic or photometric redshifts (black circles). 
Because the negative $K$-correction in the FIR/submillimeter closely offsets
the dimming effects of distance for $z>1$ (Blain et al.\ 2002; Casey et al.\ 2014), 
the observed-frame submillimeter flux is a proxy for FIR luminosity
for star-forming galaxies.  
(Note that the precise conversion of submillimeter flux to FIR luminosity 
depends on the SED of the galaxy.)
Thus, we can compare the submillimeter flux with the 20\,cm power, which should 
also measure the SFR if it is dominated by diffuse synchrotron emission produced 
by relativistic electrons accelerated in supernovae remnants. 
We denote the sources that also have SMA detections with blue circles.
Using Equation~\ref{firradio}, we also plot submillimeter flux versus radio power (blue curve)
assuming $q=2.52$ and adopting the mean conversion

\begin{equation}
\log L_{8-1000\,\mu {\rm m}}\ ({\rm erg\ s^{-1}}) = \log {S_{850\,\mu{\rm m}}\,({\rm mJy})} + 45.60\pm0.05
\label{lfir850}
\end{equation}
determined by Cowie et al.\ (2016; their Equation~3, which was based on 26 SCUBA-2 
galaxies in the GOODS-{\em Herschel\/} (Elbaz et al.\ 2011)
region with accurate positions and spectroscopic redshifts).

\vskip 0.4cm
\begin{inlinefigure}
\includegraphics[angle=0,width=3.8in]{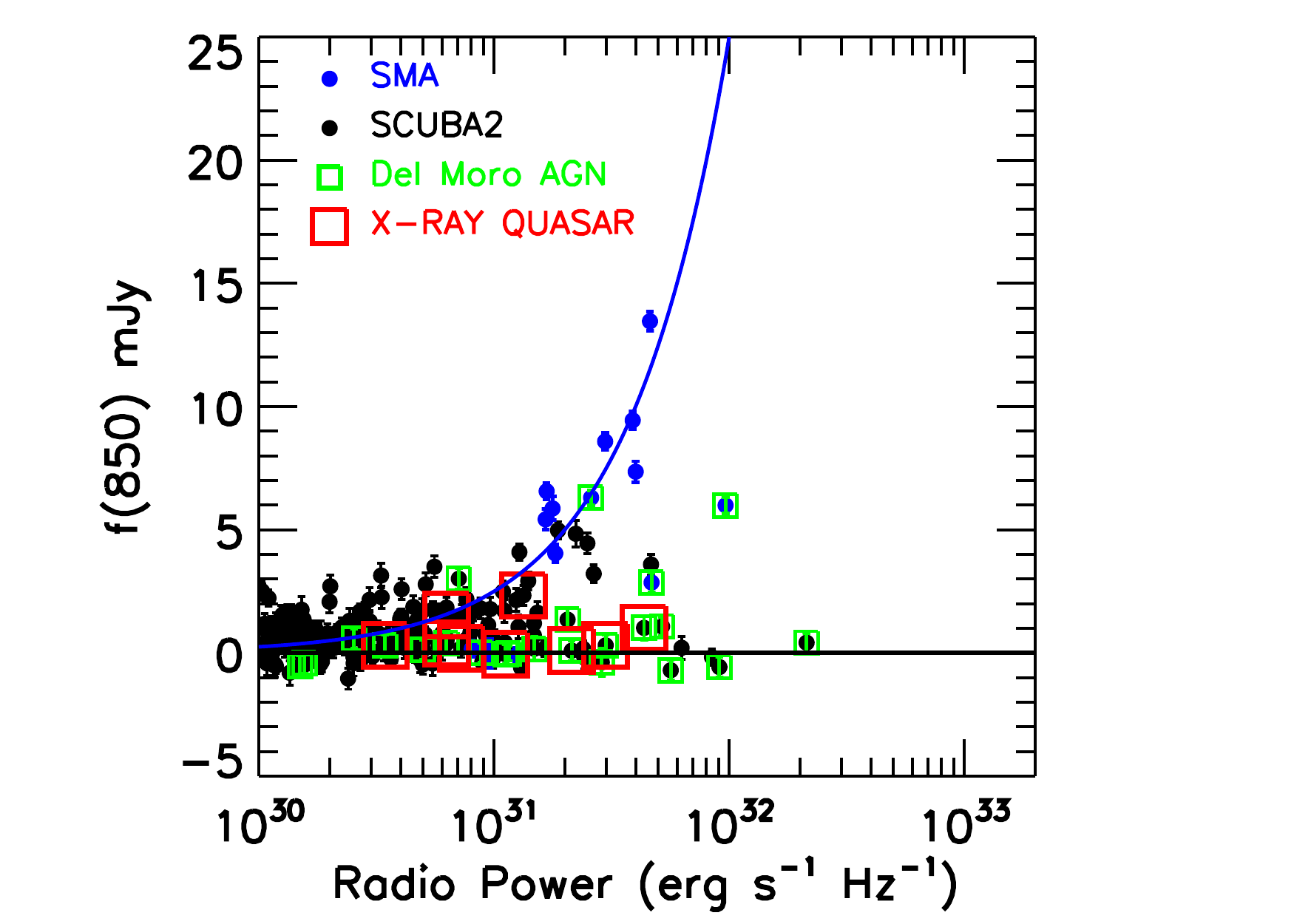}
\caption{
850~$\mu$m flux vs. 20~cm power for the radio sources 
with spectroscopic or photometric redshifts.
All submillimeter fluxes are from the SCUBA-2 image.
Sources that also have SMA detections are shown as blue circles,
while the other SCUBA-2 measurements are shown as black circles. 
All of the SMA flux measurements are consistent
with the SCUBA-2 fluxes within the errors.
Error bars are $\pm1\,\sigma$.
The blue curve shows submillimeter flux vs. radio power
based on Equations~(\ref{firradio}) and
(\ref{lfir850}) and assuming $q=2.52$. 
Sources identified as radio excess by Del Moro et al.\ (2013) 
are shown with enclosing green squares, while
X-ray quasars are shown with enclosing red squares. 
\label{barthel}
}
\end{inlinefigure}

The bifurcation between submillimeter-bright (star-forming galaxies) and submillimeter-blank 
(AGNs) is clear, with only a few sources lying in a region that may indicate they are composites
(where the radio power has contributions
from both star formation and AGN activity).
We mark with enclosing red squares the X-ray quasars in the field, all of which lie on the 
submillimeter-blank track.

Before proceeding to a comparison of our classifications with radio-excess,
VLBI, and X-ray classifications, 
it is important to note that our analysis does not include the 
radio sources without spectroscopic or photometric redshifts in the area. 
These sources are faint in the optical and NIR. Based on the $K-z$ relation,
they are likely to lie at high redshifts (Barger et al.\ 2014). 
The higher redshifts (which shifts emission lines out of the observable spectral range) 
and faintness make them hard to identify with optical and NIR spectroscopy.
However, they appear to contain a similar fraction of submillimeter galaxies as the
high radio power sources with redshifts; $34\pm9$\% of those
without redshifts have 850\,$\mu$m detections
above a $3\,\sigma$ threshold, compared to $46\pm9$\% of those with redshifts
and a radio power above $10^{31}$\,erg\,s$^{-1}$\,Hz$^{-1}$. 
Thus, while the unidentified sources may
comprise the higher redshift tail of the high radio power sample, 
the submillimeter detections suggest they may be similar to
the identified sources in other regards.

\subsection{Radio Excess}
Del Moro et al.\ (2013) looked for radio-excess candidates---those with excess
radio emission over that expected from star formation processes and hence likely to host AGN
activity---in a 24~$\mu$m detected sample
with $>3\,\sigma$ 20~cm flux measurements (their VLA/24~$\mu$m sample). 
The 20~cm fluxes were measured from the 
CDF-N radio image of Morrison et al.\ (2010; rms noise level of $\sim3.9~\mu$Jy) at the
positions of the $24~\mu$m sources (which are the same as the positions of the $3.6~\mu$m 
sources, since the IRAC 3.6~$\mu$m sources were used as priors). 
They performed a detailed SED analysis of the 458 VLA/24~$\mu$m sources with $z\le3.0$ 
and then calculated the FIR flux by integrating the total SEDs over the rest-frame wavelength range
$\lambda=42.5-122.5~\mu$m. They defined radio-excess sources as having $q<1.68$. 

We examined their sample of 51 radio-excess candidates
with our deeper VLA data and larger redshift catalog and found about one-quarter of them
to be spurious. Most of the problems seemed to come from up-scattered sources in
the noisier Morrison et al.\ (2010) 20~cm image being identified as radio-excess sources.  
The data from the upgraded VLA used in this work rule these candidates out, as well as
one candidate that is part of a radio jet.
Several other sources were allocated incorrect redshift identifications, 
including HDF850.1, which is at $z=5.124$ (Walter et al.\ 2012)
but was placed at the redshift of the neighboring elliptical galaxy. 
After removing the spurious sources, there are 37 radio-excess candidates left.  
In Figure~\ref{barthel}, we show with enclosing green squares the radio-excess
candidates that overlap with our sample. They mostly fall in the AGN regime,
as expected,
though one object is clearly a star-forming galaxy and several are composites.

\subsection{VLBI}
Next, we compare our classifications with those made from VLBI 20~cm observations.
First, in Figure~\ref{radio_ratio}, we show the information presented in Figure~\ref{barthel} 
(but restricted to sources that match our high radio power definition of 
$P_{20~\rm cm}> 10^{31}$~erg~s$^{-1}$~Hz$^{-1}$) as a ratio plot,
which allows us to illustrate the full dynamic range of the data.
We plot the ratio of $850~\mu$m flux to 20~cm power versus
20~cm power.  We show the radio sources for which the
submillimeter flux measurement at the radio position has a signal-to-noise ratio (S/N) 
$\ge2\,\sigma$ as circles with $1\,\sigma$ uncertainties, and we show
S/N$<2\,\sigma$ as circles at the $2\,\sigma$ limit with downward pointing arrows. 
We use colors to denote the sources classified as either compact (Chi et al.\ 2013;
cyan) or extended (Momjian et al.\ 2010; red) from VLBI data.
Finally, we denote with green shading the factor of two range found by 
Barger et al.\ (2014) over 
which we expect the SFRs derived from the submillimeter fluxes to be consistent with the 
SFRs derived from the radio power, assuming the FIR-radio correlation with the radio 
emission dominantly powered by star formation.
This multiplicative factor of two is the systematic error in the individual SFRs 
determined from the submillimeter fluxes based on the variations in their SEDs
(see also Cowie et al.\ 2016).

All of the radio sources within the green shaded region 
have S/N$\ge2\,\sigma$ submillimeter flux measurements, while only a handful of the radio sources 
that lie below the green shaded region do. The sources lying below the green shaded
region have submillimeter flux measurements or upper limits that are well below
what would be expected if the radio power were produced by star formation.
Thus, the radio power in these sources must be dominated by AGN contributions.
All of the VLBI compact sources lie below the green shaded region, 
while the one VLBI extended source lies within the green shaded region.

\vskip 0.5cm
\begin{inlinefigure}
\includegraphics[width=3.8in]{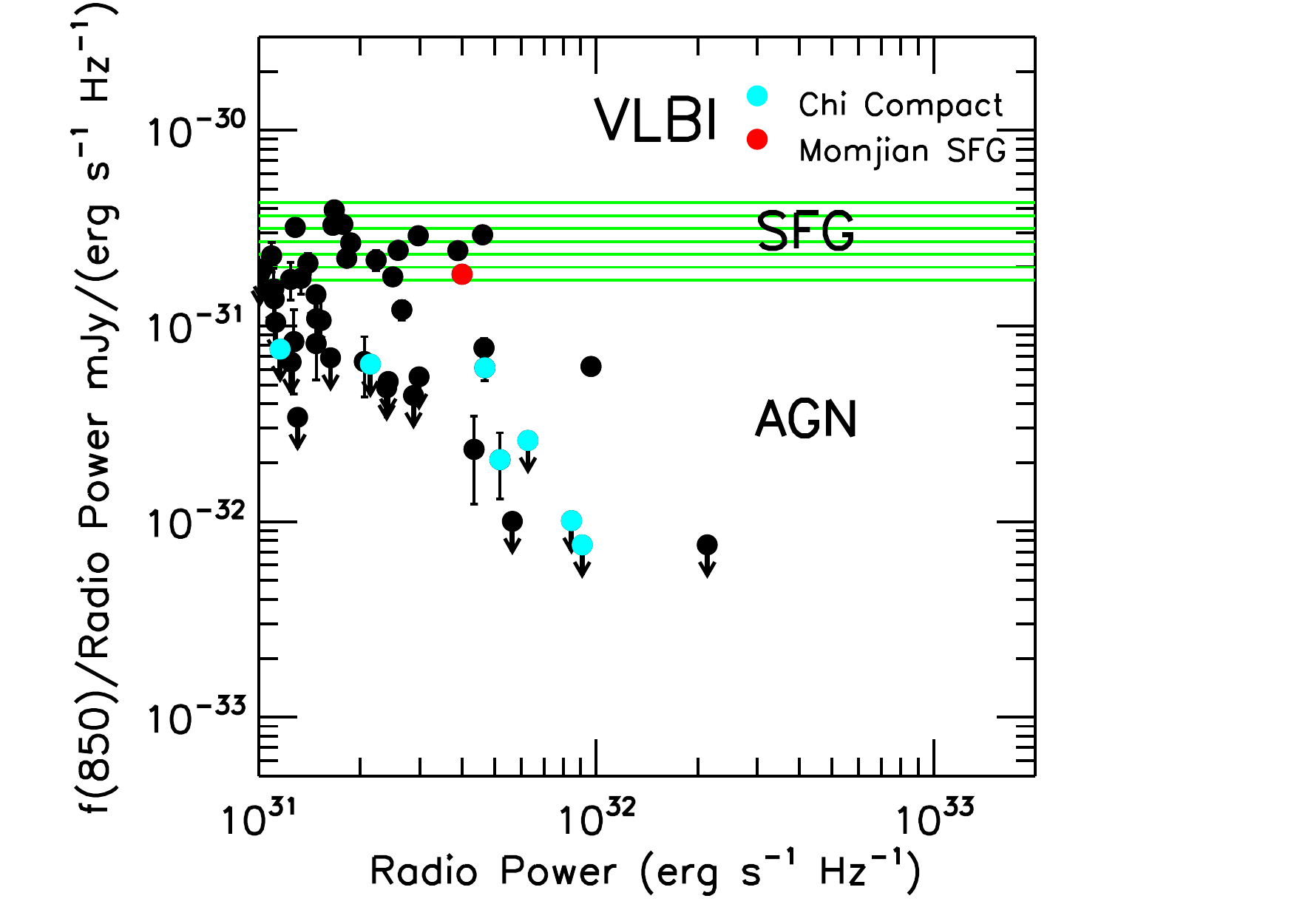}
\caption{
850~$\mu$m flux/20~cm power vs. 20~cm power for the radio sources 
with spectroscopic or photometric redshifts and
$P_{20~\rm cm}> 10^{31}$~erg~s$^{-1}$~Hz$^{-1}$ (circles).
Sources with S/N$\ge2\,\sigma$ submillimeter flux measurements are shown
with $1\,\sigma$ uncertainties, while those with S/N$<2\,\sigma$ are shown at the 
$2\,\sigma$ limit with downward pointing arrows. The green shaded region shows 
where the submillimeter flux and radio power produce consistent estimates of the 
SFRs. Cyan and red circles show sources 
classified as AGNs and star-forming galaxies (SFGs), respectively,
from VLBI observations.
\label{radio_ratio}
}
\end{inlinefigure}

Perhaps most striking, above $P_{20~\rm cm}\approx 5\times 10^{31}$~erg~s$^{-1}$~Hz$^{-1}$ there
are no sources in the green shaded region, which is consistent with the claim by Karim et al.\ (2013)
and Barger et al.\ (2014) that there is a characteristic maximum SFR for star-forming galaxies
above which the number of galaxies drops extremely rapidly. 
We note that although more luminous star-forming galaxies, such as the submillimeter galaxy GN20 
(Pope et al.\ 2005; Daddi et al.\ 2009b), which lies just outside the 124~arcmin$^2$ area studied 
here, do exist, they are rare. GN20, which  has 
$P_{20~\rm cm} = 7.5\times 10^{31}$~erg~s$^{-1}$~Hz$^{-1}$, is 
one of only two sources with  850~$\mu$m fluxes  above
15~mJy in the more extended 400~arcmin$^2$ region of the
CDF-N covered by the SCUBA-2 image to an rms of 1.5~mJy (Cowie et al.\ 2016). 
Based on the submillimeter number densities, the surface density of these
more luminous star-forming galaxies with
$P_{20~\rm cm} > 5\times 10^{31}$~erg~s$^{-1}$~Hz$^{-1}$
is about 0.01 times the surface density of star-forming galaxies with 
$P_{20~\rm cm} = (1-5)\times 10^{31}$~erg~s$^{-1}$~Hz$^{-1}$
(the $\pm1\,\sigma$ range is 0.006 to 0.037).

Hereafter, we classify the sources that lie inside the green shaded region as 
star-forming galaxies, the small number of sources that lie below that region but have
$\ge3\,\sigma$ submillimeter detections as composites, and the remaining sources as AGN dominated.

\subsection{X-ray}

\vskip 1cm
\begin{inlinefigure}
\includegraphics[width=3.1in]{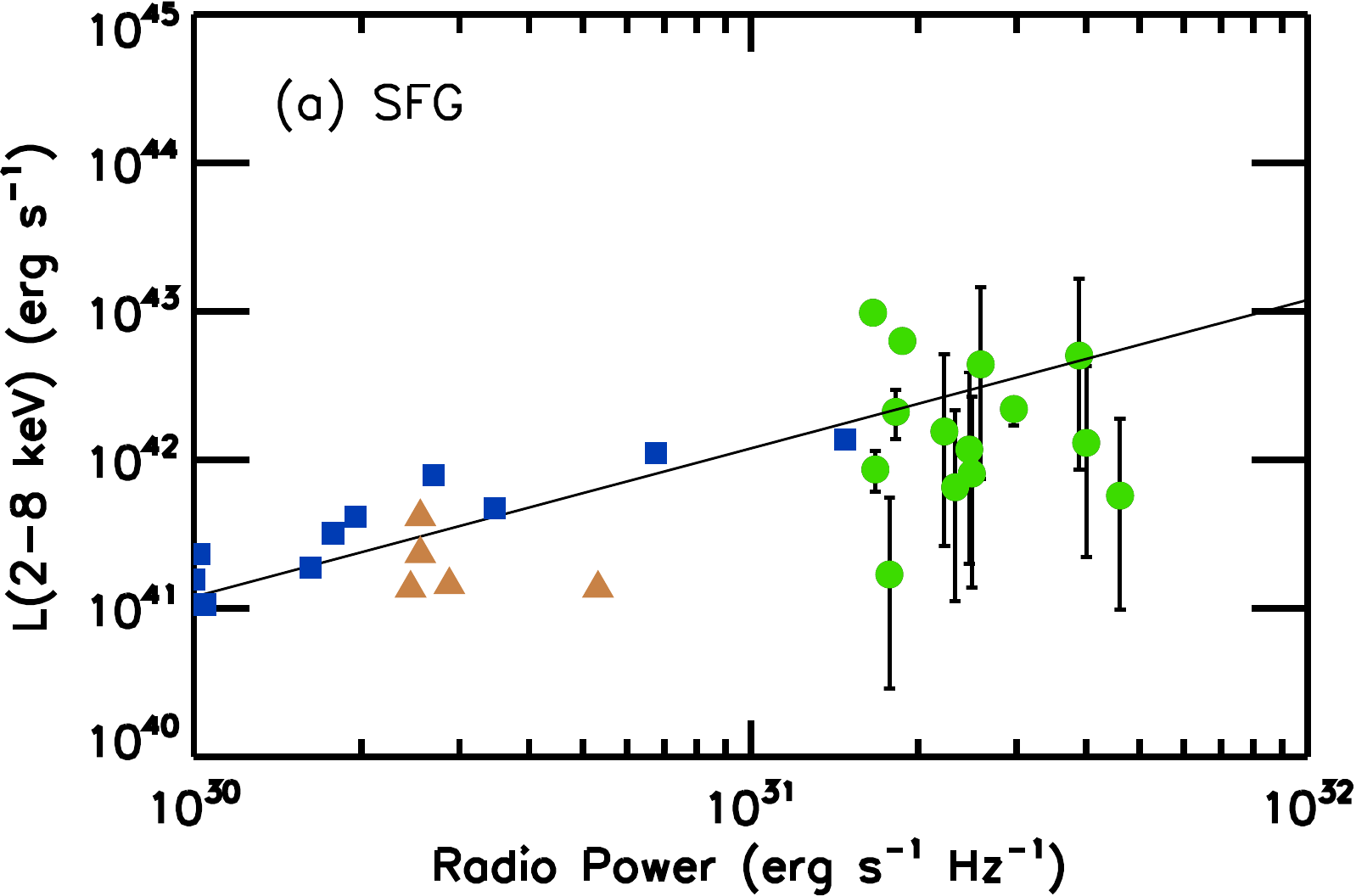}
\includegraphics[width=3.1in]{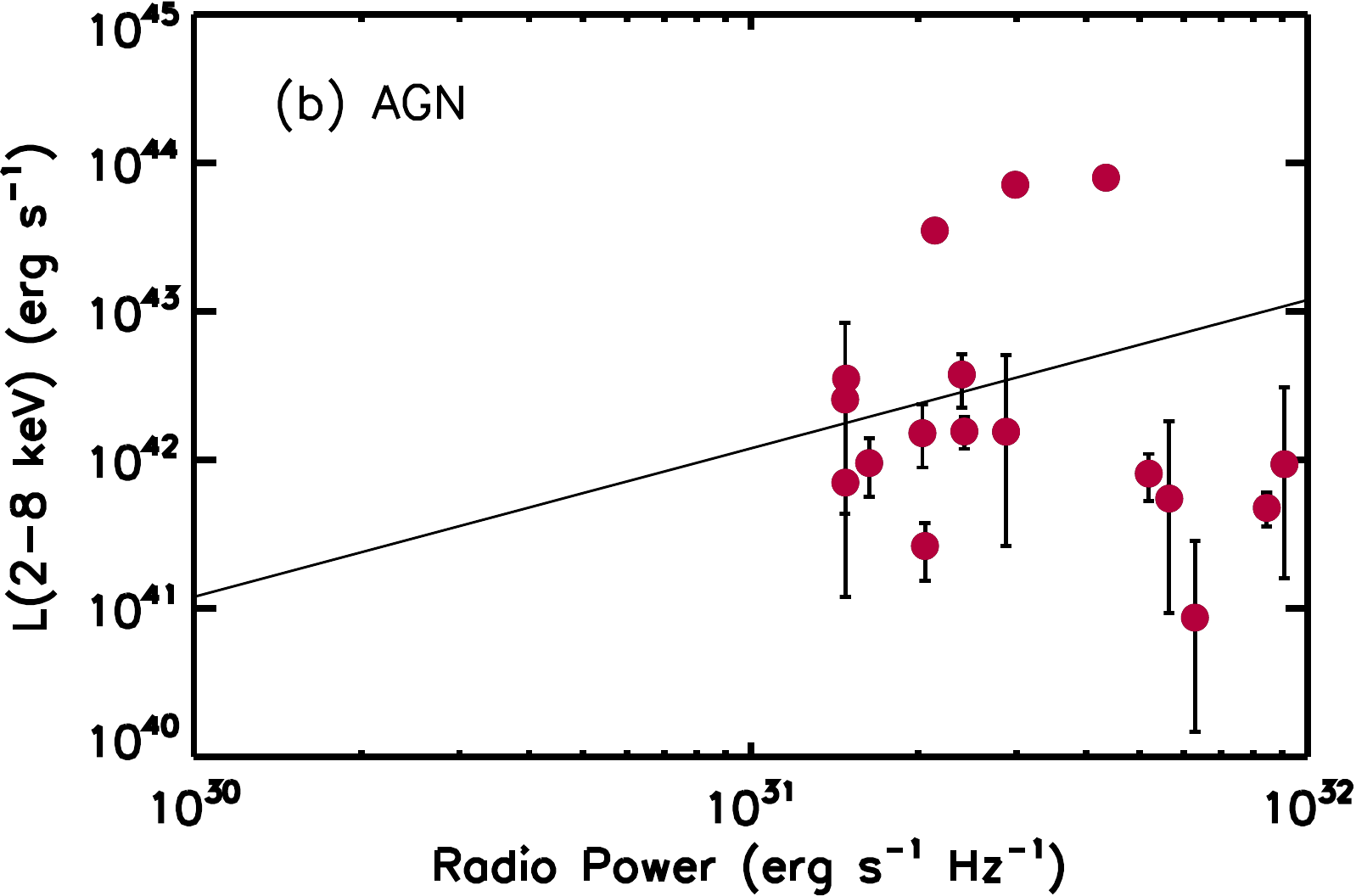}
\caption{
Rest-frame $2-8$~keV luminosity vs. radio power for the radio sources with  
spectroscopic or photometric redshifts and
$P_{20\,\rm cm} > 1.4 \times 10^{31}$~erg~s$^{-1}$~Hz$^{-1}$ and either
(a) consistent SFRs from the radio and submillimeter (green circles)
or (b) not detected in the submillimeter (red circles). 
In (a), we also show the sample of local 
ULIRGs (gold triangles) and CDF-N star-forming galaxies with 
$z=1-1.3$ (blue squares) from Mineo et al.\ (2014), together with their 
fit of the X-ray luminosity vs. radio power (black line).
\label{radpower_ls}
}
\end{inlinefigure}

We now compare our classifications with the X-ray
properties of the sample. In Figure~\ref{radpower_ls}, we show
rest-frame $2-8$~keV luminosity (computed from the observed-frame 
$0.5-2$~keV flux, as described in Section~\ref{xraydata},
to take advantage of the higher sensitivity in this energy band) 
versus the 20~cm power for the radio sources with
spectroscopic or photometric redshifts and
$P_{20\,\rm cm}>1.4\times 10^{31}$~erg~s$^{-1}$~Hz$^{-1}$.
The use of the $0.5-2$~keV flux to compute the $2-8$~keV luminosity 
also minimizes any uncertainty in the $K-$corrections computed from the
X-ray spectral slope with the relation being exact at $z=3$.
We chose this radio power threshold to provide a clean separation
in Figure~\ref{radio_ratio} between star-forming galaxies and AGNs.

At lower redshifts ($z<1.3$) and radio powers, Mineo et al.\ (2014)
compared X-ray luminosity with radio
power for a sample of galaxies from the CDF fields that they
considered to be star-formation dominated based on their colors
and morphologies. In Figure~\ref{radpower_ls}(a),
we show their CDF-N sample with 
$P_{20~\rm cm} > 10^{30}$~erg~s$^{-1}$~Hz$^{-1}$
(blue squares) after converting their 
$0.5-8$~keV luminosities to $2-8$~keV luminosities using a factor 
of 0.56. We also show their measurements for a sample of local
ULIRGs (golden triangles).
Mineo et al.\ fitted their data over four orders of magnitude
in radio power with the linear relation 
\begin{equation}
L_{2-8~{\rm keV}}\ ({\rm erg~s^{-1})} = 1.2\pm0.1 \times 10^{11} {P_{\rm 20\,cm}~({\rm erg~Hz^{-1}})}
\label{mineo}
\end{equation}
(black line in Figure~\ref{radpower_ls}). Lehmer et
al.\ (2016) discuss the dependence of the Mineo et al.\ relation on the specific
SFRs, but for very high specific SFRs, they reproduce
the Mineo et al.\ equation, which should be appropriate for the present objects.

The X-ray luminosities of the star-forming galaxies in Figure~\ref{radpower_ls}(a)
are consistent with an extrapolation of the Mineo et al.\ (2014) equation
to higher radio powers. The mean
$2-8$~keV luminosity to radio power ratio of our
star-forming galaxies is $1.22\pm0.43 \times 10^{11}$~Hz,
where we have computed the error using the jackknife resampling statistic. 
This is almost identical to the Mineo et al.\ normalization, extending the
applicability of their relation
to galaxies with SFRs $>1000~M_\odot~{\rm yr}^{-1}$. 
The extremely high SFRs of the present sample result in some of the X-ray 
luminosities being even greater than the $L_{2-8~{\rm keV}} = 10^{42}$~erg~s$^{-1}$ 
value often used to identify sources as clear AGNs
(e.g., Hornschemeier et al. 2001; Barger et al.\ 2002; Bauer et al.\ 2004;
Szokoly et al.\ 2004) based on maximal local star-forming galaxy X-ray luminosities.
Laird et al.\ (2010) similarly concluded that very high SFRs were responsible for 
the high X-ray luminosities of most of the sources in their submillimeter-selected 
sample in the same field.
 
The AGN-classified sources in Figure~\ref{radpower_ls}(b)
show a much wider spread in $L_{2-8~{\rm keV}}$,
ranging from near quasar X-ray luminosities to nearly X-ray undetected sources.
However, there are a number of sources with luminosities
close to those of the star-forming galaxies, and
there is clearly not a one-to-one correlation between our 
classifications and X-ray luminosity.

\section{The FIR-Radio Correlation}
\label{seccorr}
Radio power is often used along with the FIR-radio correlation as a measure 
of the SFR (e.g., Cram et al.\ 1998; Hopkins et al.\ 2003; Netzer et al.\ 2007; 
Mushotzky et al.\ 2014; Mineo et al.\ 2014).
However, this could be risky when no other information is available 
on whether a source is AGN dominated or star-formation dominated. After all,
it is well known that some local radio-quiet AGNs lie on the FIR-radio 
correlation (e.g., Condon 1992; Mori{\'c} et al.\ 2010; Wong et al.\ 2016), 
even if it is not understood why.

\vskip 0.5cm
\begin{inlinefigure}
\includegraphics[width=3.0in]{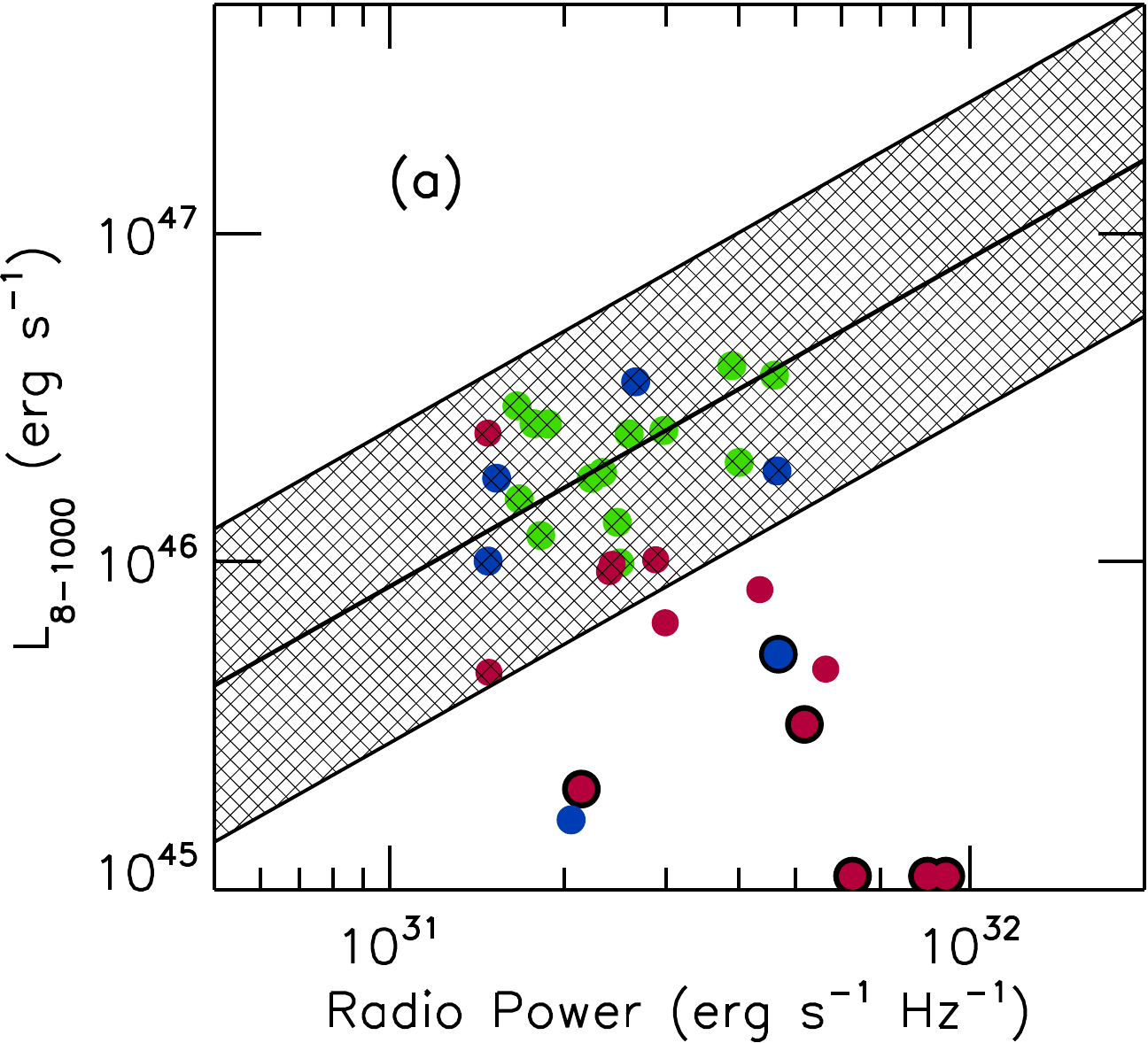}
\includegraphics[width=3.0in]{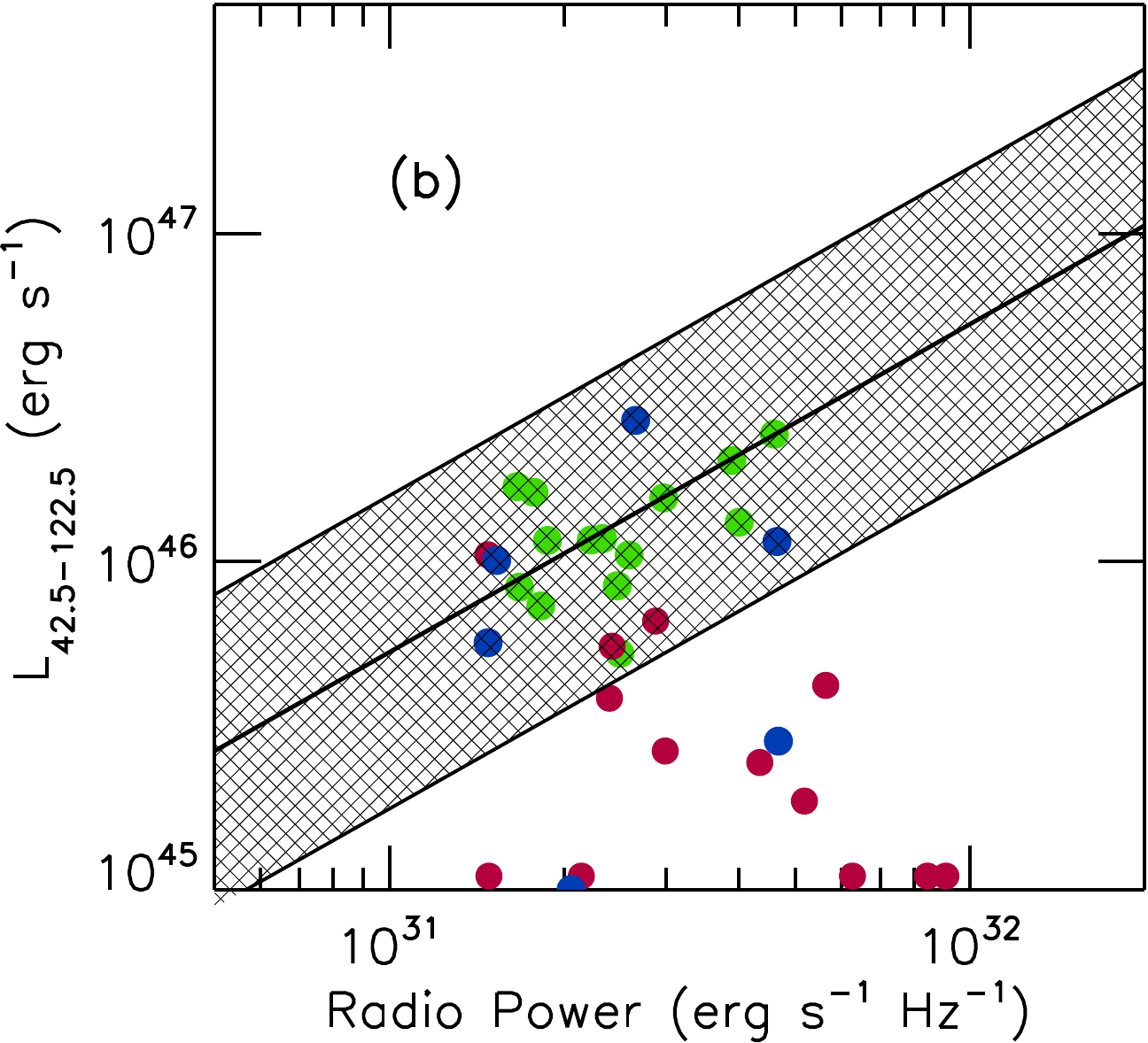}
\caption{
(a) $8-1000\,\mu {\rm m}$ luminosity and
(b) $42.5-122.5\,\mu {\rm m}$ luminosity vs. 20\,cm power for the radio sources 
with spectroscopic or photometric redshifts and
$P_{20\,\rm cm}>1.4\times 10^{31}$~erg~s$^{-1}$~Hz$^{-1}$.
Green circles denote star-forming galaxies, blue circles composites, 
and red circles AGNs. The central black line shows the median FIR-radio 
correlation of the star-forming galaxies, and the shading shows
the region where we consider the sources to lie on the FIR-radio
correlation (multiplicative factor of 3 of the central value).
In (a), we mark the sources that Chi et al.\ (2013) found to
be AGNs using VLBI observations with surrounding black circles. 
All of these sources lie below the FIR-radio correlation.
\label{correl}
}
\end{inlinefigure}

In Figure~\ref{correl}, we plot (a) $8-1000\,\mu$m luminosity 
and (b) $42.5-122.5\,\mu {\rm m}$ luminosity versus radio 
power for the radio sources with spectroscopic or photometric redshifts 
and $P_{20\,\rm cm}>1.4\times 10^{31}$~erg~s$^{-1}$~Hz$^{-1}$.
We use colors to denote the sources classified as star-forming galaxies
(green circles), composites (blue circles), and AGNs (red circles).
In (a), we also mark the sources that Chi et al.\ (2013) found to
be AGNs using VLBI observations with surrounding black circles. 
All of these sources lie below the FIR-radio correlation.
The star-forming galaxies in (a) have a mean 
$q=2.42\pm0.06$, where we have
computed the error with the jackknife resampling statistic;
all lie within a multiplicative factor of three of the median 
value ($q=2.35$), which we denote by shading.
Five AGNs and four composites also lie in this region.

\vskip 0.5cm
\begin{inlinefigure}
\includegraphics[width=3.4in]{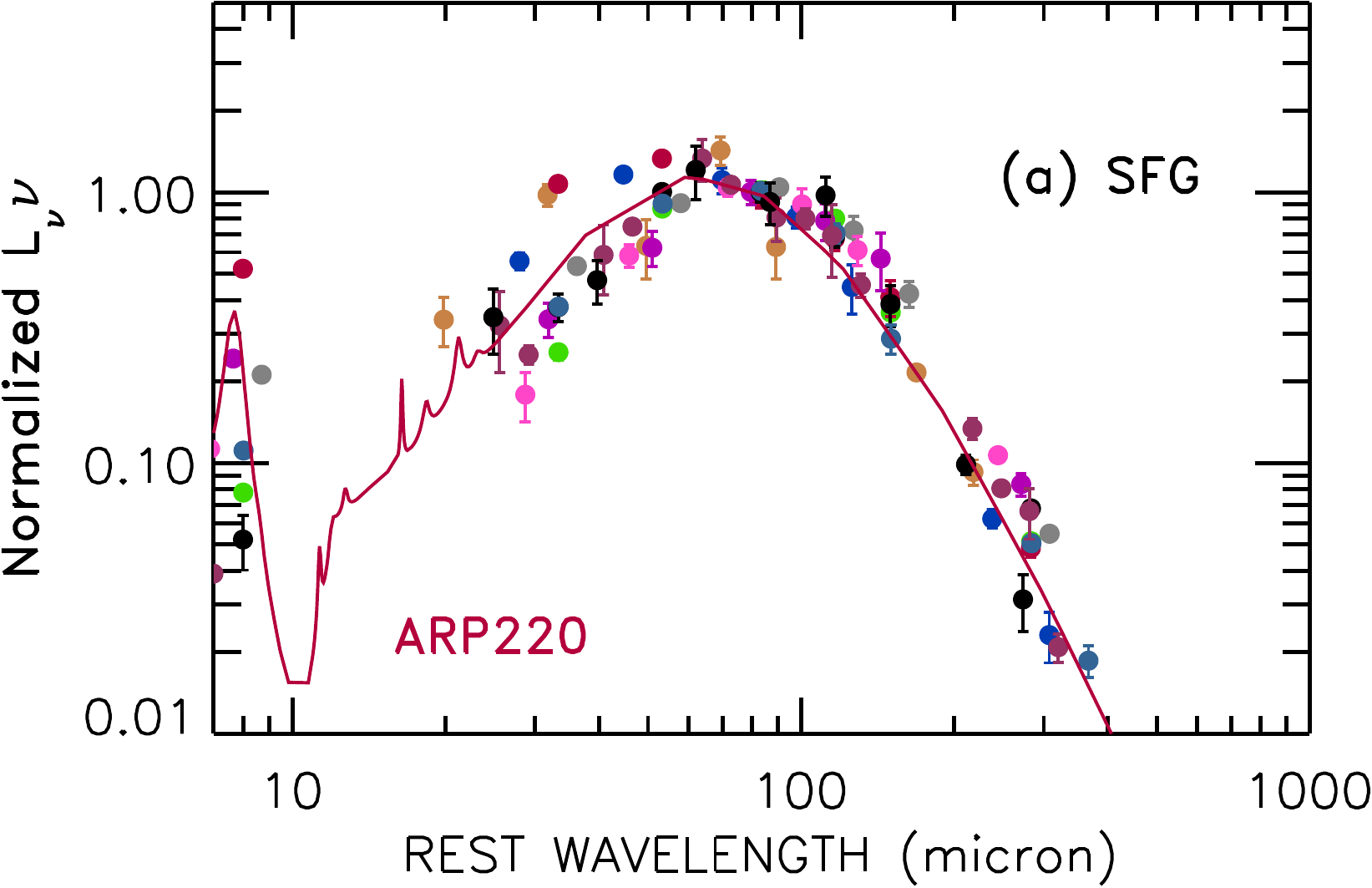}
\vskip 0.5cm
\includegraphics[width=3.4in]{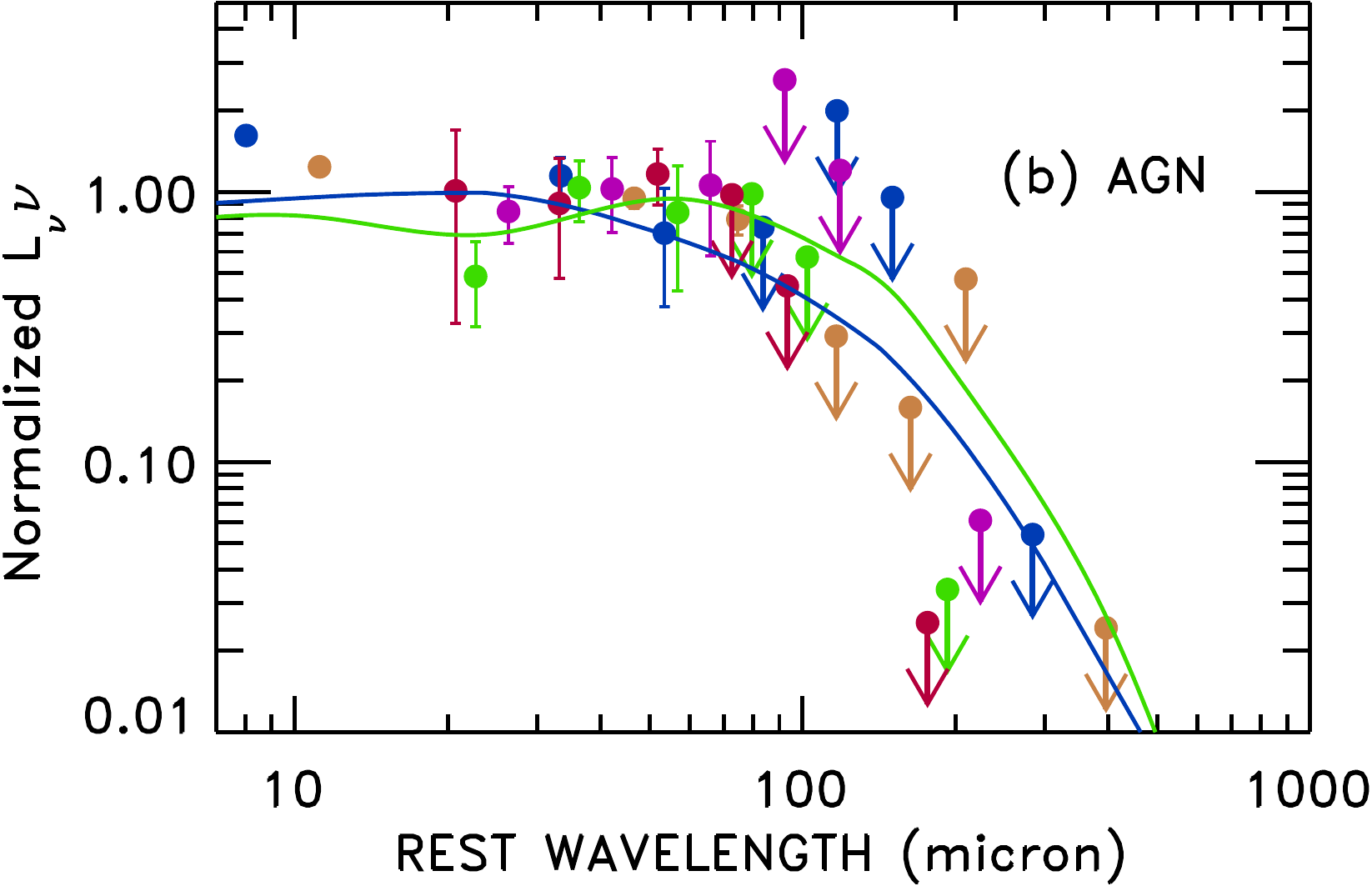}
\caption{SEDs for the radio sources with spectroscopic or 
photometric redshifts and $P_{20~\rm cm}> 1.4\times 10^{31}$~erg~s$^{-1}$~Hz$^{-1}$ 
that lie on the FIR-radio correlation (i.e., the shaded region in Figure~\ref{correl}):
(a) 12 star-forming galaxies (excluding the highest redshift
HDF850.1 where the short wavelengths are poorly
defined), and (b) 5 AGNs. The error bars are $\pm1\,\sigma$. 
When a source is not detected at the $2\,\sigma$
level at a particular wavelength, we show it at the $2\,\sigma$ value 
with a downward pointing arrow. The star-forming galaxies in (a) show the
characteristic shape of luminous starburst galaxies, which we illustrate
using the Arp~220 SED from Silva et al.\ (1998) (red curve). 
The AGNs in (b) show a torus-dominated 
SED, which we illustrate using the SWIRE templates for type~I 
quasars (blue curve) and type~II quasars (green curve). 
\label{SEDs}
}
\end{inlinefigure}

In Figure~\ref{SEDs}(a), we show the SEDs of 12 of the star-forming galaxies
that lie on the FIR-radio correlation, as defined by
the shaded region of Figure~\ref{correl}(a). We excluded the 
highest redshift source HDF850.1, because the short wavelength data are not well 
defined. These star-forming galaxies show the characteristic
shape of luminous starburst galaxies, as we illustrate by superimposing the Arp~220 
SED from Silva et al.\ (1998) (red curve).

In Figure~\ref{SEDs}(b), we show the five AGNs from Figure~\ref{correl}(a)
that lie on the FIR-radio correlation.
They show typical torus SEDs and are undetected at wavelengths 
$\gtrsim60$\,$\mu$m in the rest frame. We superimpose the
quasar type~I (blue curve) and type~II (green curve) templates from the SWIRE template
library\footnote{\url{http://www.iasf-milano.inaf.it/$\sim$polletta/templates/swire templates.html}}
(Polletta et al.\ 2006, 2007). Symeonidis et al.\ (2016) 
give a very similar SED for the type~I quasars based on a carefully
selected and analyzed sample. 

The template quasar SEDs are broadly
similar to our observations, though our observed SEDs appear to
fall more sharply at the longer wavelengths. For these AGNs,
the FIR luminosity and the radio power are not a measure of the
SFR in the host galaxies. Indeed, the
fact that these AGNs lie on the FIR-radio correlation seems to be a 
coincidence. From Figure~\ref{correl}(b), we see that when we
use the $42.5-122.5\,\mu$m luminosity instead of the $8-1000\,\mu$m 
luminosity, the locations of the star-forming galaxies
remain about the same, while the AGNs drop substantially.

\section{Radio Sizes}
\label{secsizes}
In previous work, radio morphologies and/or radio spectral indices
have been used, together with mid-infrared (MIR) data, to pick out star-forming 
galaxies and then measure their radio sizes 
(Muxlow et al.\ 2005; Guidetti et al.\ 2013). Muxlow et al. found
sizes from $0\farcs2$--$3\farcs0$ with a median size of $1\farcs0$. 
They also found that most of the AGNs were compact.
In other work, radio sizes of submillimeter and millimeter 
selected galaxies have been measured 
(Chapman et al.\ 2004; Biggs \& Ivison 2008; Miettinen et al.\ 2015).
The advantage of the present data is that we can compare the radio sizes of the 
classified star-forming galaxies and AGNs in our
purely radio selected sample without
any selection dependence on the radio morphologies.
We could, in principle, use the submillimeter sizes from the SMA 
data to carry out a similar exercise, but the resolution and 
signal-to-noise is too low for this to be useful.

In Figure~\ref{radio_sizes}, we show radio size versus the 
ratio of 850\,$\mu$m flux 
to radio power for the $f_{20\,\rm cm}>40\,\mu$Jy sources with 
spectroscopic or photometric redshifts
and $P_{20\,\rm cm}>1.4\times 10^{31}$\,erg\,s$^{-1}$\,Hz$^{-1}$.
We apply the radio flux limit to ensure high enough 
S/Ns for accurate deconvolution to be able to measure
the radio sizes. This limit was based on model tests within the 
current data set.
We show the unresolved sources with downward pointing arrows.
We again use colors to denote the sources classified as 
star-forming galaxies (green), composites (blue), or AGNs (red). We use larger black
circles to highlight the two double-lobed radio sources in our sample, which
are very extended in the radio.

All but one of the star-forming galaxies are resolved in
the 20\,cm image, and all but one of the resolved star-forming
galaxies have sizes in
the rather narrow range $0\farcs42$ to $1\farcs36$ (physical
sizes from 4~kpc to 12~kpc). The one larger source
is the heavily studied source SMG123707+621408.
This source has two separated
components, each of which is individually similar to the
remaining star-forming galaxies. However, since the CO map of 
Tacconi et al.\ (2006) appears to join the two components, we have 
used the full extent of the two components in measuring the radio
size rather than splitting them. We note, however, that the 
submillimeter continuum flux arises from only one of the components. 
Deciding what size to allocate to such sources is difficult;
fortunately, though, such sources are not common, with only this one 
example lying in our observed region.

We show the distributions of the sizes in the AGN (top)
and star-forming galaxy (bottom) populations in Figure~\ref{radio_hist}, 
where we have put the unresolved sources in at the upper limits on their 
sizes (open portions of histograms).
The median size of the star-forming galaxies is $1\farcs0\pm0.3$, while that of
the AGNs (excluding the double-lobed sources) is $0\farcs45\pm0.05$.
(The value of the AGNs drops to $0\farcs4$
if we put the unresolved sources in at zero.) 

There are only two composites, one of which is unresolved.
If we conservatively put the three unresolved AGNs at their upper limits
and include both composites in a combined AGN/composite 
category (putting the unresolved composite at its upper limit), 
then the differences in the size distributions between the star-forming
galaxies and the AGNs/composites are highly significant, with
a Mann-Whitney test giving only a 0.002 probability (two sided)
that the two distributions are consistent. 

Even with the current 20\,cm resolution, we can pick out
most of the star-forming galaxies based on radio size and morphology. 
The double-lobed radio sources are easily classified as AGNs on the 
basis of radio morphology, while for the remaining sources,
we find that choosing those that are resolved and greater than 
$0\farcs5$ (blue dotted line in Figure~\ref{radio_sizes})
would correctly classify the bulk of the star-forming galaxies, while 
avoiding contamination from the resolved AGNs. However, it should be
noted that the most compact
star-forming galaxies have sizes that are hard to distinguish from the 
AGNs using the present spatial resolution.

\vskip 1cm
\begin{inlinefigure}
\includegraphics[width=3.4in]{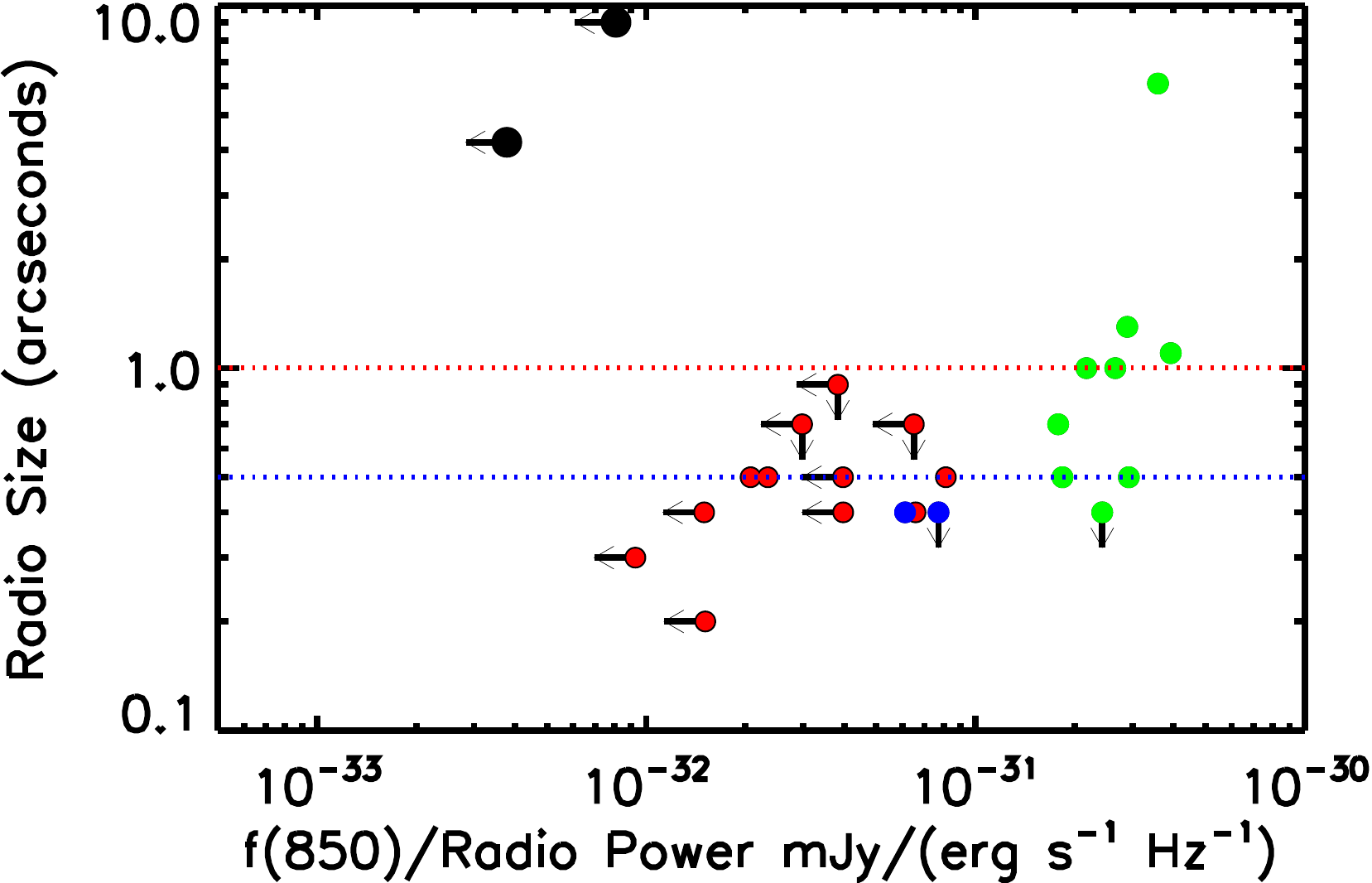}
\caption{
20\,cm size vs. the ratio of 850\,$\mu$m flux to radio power
for the $f_{20\,\rm cm}>40\,\mu$Jy sources 
with spectroscopic or photometric redshifts and
$P_{20\,\rm cm}>1.4\times 10^{31}$~erg~s$^{-1}$ Hz$^{-1}$.
Unresolved sources are shown with downward pointing arrows. 
Green circles denote star-forming galaxies, blue circles composites, 
and red circles AGNs. 
The larger black circles show the double-lobed radio sources 
in the field both classified as AGNs based on their radio morphology; 
these are very extended in the radio. 
The red dotted line shows the median size of the star-forming galaxies.
The blue dotted line shows a dividing line of $0\farcs5$ that would
correctly classify most of the resolved sources.
\label{radio_sizes}
}
\end{inlinefigure}

\vskip 0.4cm
\begin{inlinefigure}
\includegraphics[width=3.1in]{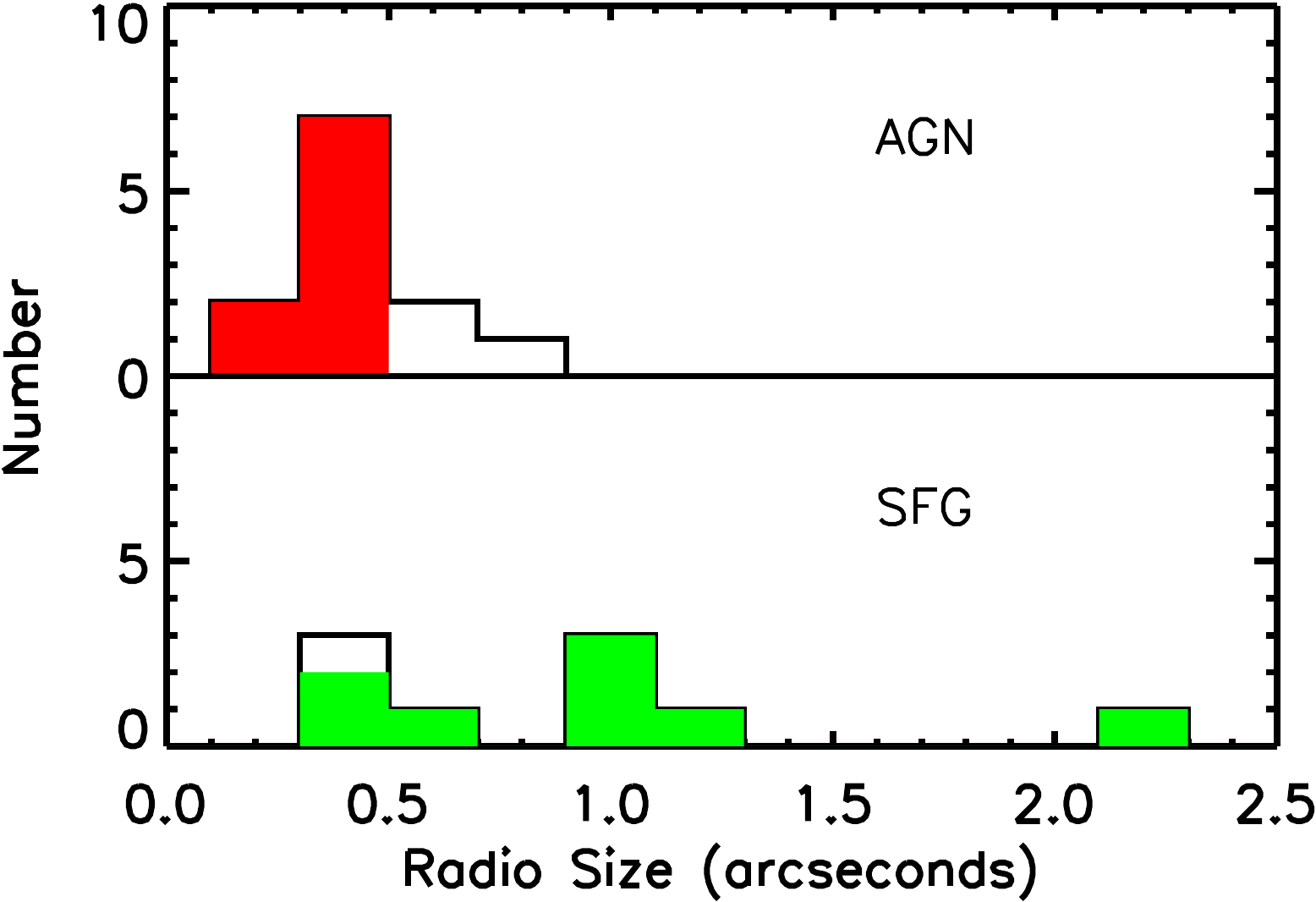}
\caption{
Histogram of 20\,cm size for the $f_{20\,\rm cm}>40\,\mu$Jy
sources with spectroscopic or photometric redshifts and
$P_{20\,\rm cm}>1.4\times 10^{31}$~erg~s$^{-1}$ Hz$^{-1}$.
These are separated into AGNs (top) and star-forming galaxies (bottom).
The two double-lobed radio sources, which have large sizes, are not shown
in the AGN panel. In the star-forming galaxy panel, the source SMG123707+621408
discussed in the text is shown at a nominal value of $2\farcs2$ to keep
it on the plot.
The colored histograms show sources with resolved sizes, and the open
histograms show sources that are not resolved, put at the upper limits on
their sizes. These sources could lie anywhere to the left.
\label{radio_hist}
}
\end{inlinefigure}

The radio sizes for the star-forming galaxies are comparable to
(though slightly larger than) the radio sizes measured for samples 
of galaxies with submillimeter detections
by Chapman et al.\ (2004) ($0\farcs83\pm0.14$) and
Biggs \& Ivison (2008) ($0\farcs64\pm0.14$). 
However, all of these radio sizes are generally larger than those 
measured from submillimeter continuum observations
(typically about $0\farcs3$; Simpson et al.\ 2015) by 
factors of $2-3$. It is unclear what the explanation for this is when we 
have found here that the radio power is a good tracer of the FIR for the 
high radio power selected star-forming galaxy sample 
(see Figure~\ref{correl}). Simpson et al.\ suggest that
it may be a consequence of cosmic ray diffusion, but they then 
argue that this is unlikely for the strong submillimeter galaxies. 
(See also Miettinen et al.\ 2015's Section~6.3 for a discussion of 
this issue based on a comparison of their 10\,cm sizes with the 
millimeter continuum sizes from Ikarashi et al.\ (2015).) 
The origin of the effect therefore remains to be determined.

\section{Summary}
\label{disc}
In this paper, we used ultradeep radio and submillimeter
data on the GOODS-N/CDF-N to develop a classification method 
based on the ratio of the submillimeter flux to the radio power
to separate star-forming galaxies from AGNs in the high radio 
power population. We found that our classifications agreed with classifications 
made from high-resolution VLBI observations. However,
we also found that there was not a one-to-one correlation of our classifications 
with those made from ultradeep X-ray data on the field.
The star-forming galaxies agreed well with an extrapolation of a local
relation between X-ray luminosity and radio power.
In contrast, the AGNs showed a much
wider spread in hard X-ray luminosity, including some that had luminosities
close to those of the star-forming galaxies.
These results suggest that some high-redshift, extremely high star formation 
sources may have been incorrectly classified as AGNs if a simple $2-8$~keV 
luminosity threshold of $10^{42}$~erg~s$^{-1}$ was used. 

We examined the FIR-radio correlation for the classified sources and found that
the star-forming galaxies were all within a multiplicative factor of three of the
median FIR-radio correlation measured from the star-forming galaxies. However,
5 AGNs also lay within this region. We constructed the SEDs for both the
star-forming galaxies and these five AGNs and found that the star-forming
galaxies show the characteristic shape of luminous starburst galaxies, while the
AGNs show typical torus SEDs and are undetected at wavelengths
$\gtrsim60\,\mu$m. For these AGNs, we are not measuring SFR from either
the FIR luminosity or the radio power, and the fact that they lie on the FIR-radio
correlation seems to be mere coincidence. Thus, at least at these high
radio powers, one must be cautious in
measuring SFRs for radio sources using the FIR-radio correlation without other 
information available on whether they are AGN dominated or star-formation dominated.

We find that the number of star-forming galaxies drops rapidly above a radio 
power of $P_{20\,\rm cm}\approx5\times 10^{31}$\,erg\,s$^{-1}$\,Hz$^{-1}$.
The surface density of these
$P_{20~\rm cm} > 5\times 10^{31}$~erg~s$^{-1}$~Hz$^{-1}$ sources
is about two orders of magnitude lower than that
of star-forming galaxies with $P_{20~\rm cm} = (1-5) \times 10^{31}$~erg~s$^{-1}$~Hz$^{-1}$.
The $5\times 10^{31}$\,erg\,s$^{-1}$\,Hz$^{-1}$  bound corresponds to a SFR 
of just over a thousand solar masses per year using the Murphy et al.\ (2011)  
conversion for a Kroupa (2001) initial mass function.

Finally, since obtaining wide-field, deep submillimeter images is not easy, we 
used our classifications to investigate
whether radio sources could be separated into star-forming 
galaxies and AGNs based on radio sizes alone.  We found that even with the
current 20\,cm resolution, we were able to put most of the sources into the correct
class using a size of $0\farcs5$ as the separation point (i.e., AGNs were smaller than this, 
and star-forming galaxies were larger). The radio sizes of our submillimeter-detected 
radio sources are generally larger than those measured for submillimeter-selected galaxies 
from submillimeter continuum data, but the explanation for this is not yet clear.

\acknowledgements
We thank the anonymous referee for a careful report that helped us to 
improve the manuscript.
We thank P. Barthel for discussions on the VLBI results.
We gratefully acknowledge support from
NSF grants AST-1313309 (L.~L.~C.) and AST-1313150 (A.~J.~B.), and
from the John Simon Memorial Guggenheim Foundation and the Trustees
of the William F. Vilas Estate (A.~J.~B.).
The authors wish to recognize and acknowledge the very significant 
cultural role and reverence that the summit of Maunakea has always 
had within the indigenous Hawaiian community. We are most fortunate 
to have the opportunity to conduct observations from this mountain.


\end{document}